\documentclass{aa}  
\usepackage{graphicx}
\usepackage{txfonts}
\usepackage{epsfig}
\usepackage[below]{placeins}
\usepackage{6650natbib}
\usepackage{6650journalshortcuts}
\bibpunct{(}{)}{;}{a}{}{,}

\begin{document}
   \title{The gaseous atmosphere of M87 seen with XMM-Newton}


   \author{A. Simionescu \inst{1}
          \and H. B\"ohringer \inst{1}
	  \and M. Br\"uggen \inst{2}
	  \and A. Finoguenov \inst{1,3}
          }

   \offprints{A. Simionescu, aurora@mpe.mpg.de}

   \institute{Max Planck Institute for Extraterrestial Physics,
              Giessenbachstr, 85748 Garching, Germany
         \and
             International University Bremen,
	     Campus Ring 1, 28759 Bremen, Germany
	 \and
	     University of Maryland, Baltimore County, 1000
             Hilltop Circle,  Baltimore, MD 21250, USA
             }

   \date{Received 27 October 2006 / Accepted 29 December 2006}


  \abstract
   {}
   {M87 is a key object whose study can reveal the complex phenomena in cooling cores. We use a deep XMM-Newton observation of M87 to produce detailed temperature, pressure and entropy maps in order to analyze the physical processes of cooling cores and of their heating mechanisms.}
   {We employed both broad-band fitting and full spectroscopical one-temperature model analysis to derive temperature and surface brightness maps, from which the pseudo-deprojected entropy and pressure were calculated. We discuss possible physical interpretations of small deviations from radial and elliptical symmetry in these maps.}
   {The most prominent features observed are the E and SW X-ray arms that coincide with powerful radio lobes, a weak shock at a radius of 3$^\prime$, an overall ellipticity in the pressure map and a NW/SE asymmetry in the entropy map which we associate with the motion of the galaxy towards the NW. For the first time we find evidence that cold, metal-rich gas is being transported out of the center, possibly through bubble-induced mixing. Several edges in the abundance map indicate an oscillation of the galaxy along the NW/SE direction. Furthermore, the radio lobes appear to rise along the short axis of the elliptical pressure distribution, following the steepest gradient of the gravitational potential, and seem to contain a nonthermal pressure component.}
   {}


   \maketitle
%

\section{Introduction}
As early as 1973, the first detailed X-ray observations with the Uhuru and Copernicus satellites revealed that the X-ray luminosity at the center of some galaxy clusters and elliptical galaxies has a very sharp peak, implying a high enough central gas density that the cooling time falls below the Hubble time \citep{Lea73}. Consequently, the so-called cooling flow model was developed which asserted that the gas, in losing energy by means of radiation and in the absence of other heat sources, cools down to temperatures so low as to become undetectable in X-rays, and sinks towards the cluster core (e.g. \cite{FabianNulsen77}, \cite{CowieBinney77}). The model however failed to explain the ultimate fate of the cooling gas, and the evidence in wavebands other than X-rays showed that much less gas was condensing into a cooler state than the "cooling flow" model predicted (e.g. \cite{McNamara89}).

The latest generation of X-ray satellites, in particular XMM-Newton and Chandra, provided more detailed spectral information showing that indeed much less gas undergoes such a "cooling flow" than the initial model based on the X-ray luminosity peak predicted (\cite{Peterson01}, \cite{Peterson03}, \cite{Boehringer01}, \cite{Matsushita02}). This led to the conclusion that a heating source must exist which interacts with the gas in a fine-tuned way, stopping the cooling at just the right level to reproduce the observations (e.g. \cite{Boehringer02}). Active galactic nuclei (AGN) are found in most clusters previously believed to host a "cooling flow" \citep{Burns90} and represent the most favoured candidates for providing such heat input. The main challenge remains the understanding of the exact mechanisms by which the heating takes place (e.g. \cite{David01}, \cite{Churazov01}, \cite{Fabian03}, \cite{Bruggen05}, \cite{Heinz06}).

M87, the central dominant galaxy in the nearby Virgo cluster, is an ideal candidate whose study may shed light on the complex phenomena in cooling cores. M87 was recognized early on as having a peaked X-ray surface brightness corresponding, in the absence of heating, to a cooling flow with a mass deposition rate of 10 $\rm{M}_{\odot}$/year \citep{Stewart84}. However, using the spectral information from XMM-Newton, \cite{Boehringer01} put an upper limit on the mass deposition rate at least one order of magnitude below this value (see also \cite{Matsushita02}, \cite{Molendi02}). The energy input which may prevent the cooling of the gas could be supplied by the central AGN, powered by the galaxy's central supermassive black hole which has a mass of $3.2 \times 10^9 \: \rm{M}_{\odot}$ \citep{Harms94}. Based on the observed complex system of radio lobes, believed to be associated with the AGN jet and unseen counterjet, \cite{OEK00} showed that the mechanical power input from this AGN is more than sufficient to compensate for the energy loss through X-ray radiation.

The radio plasma in M87 clearly interacts with the X-ray gas. The best proof of this is the fact that the two most prominent radio lobes extending to the east and south-west of the M87 center correspond to regions of enhanced X-ray surface brightness, also known as the E and SW X-ray arms. Moreover, these E and SW arms show a more complex multi-phase temperature structure \citep{Molendi02} while a previous observation with XMM-Newton showed that the rest of the ICM is locally well approximated as single-phase \citep{Matsushita02}. This motivated \cite{Churazov01} to develop a model according to which the bubbles seen in the radio map buoyantly rise through the hot plasma uplifting cooler gas from the central region.

Further features in the X-ray surface brightness map of M87 described by \cite{Forman05} from Chandra observations include a ring of enhanced emission with a radius of 14 kpc, possibly associated with a weak shock which may be one of the mechanisms contributing to the heating of the ICM, additional arc-like features at 17 and 37 kpc, and excess emission regions NW and SE of the core. Using deeper Chandra data, \cite{Forman06} confirmed the presence of a weak shock at 14 kpc and determined its Mach number as M$\sim$1.2.

Detailed temperature, entropy and pressure maps are a prerequisite for the physical understanding of the phenomena behind these features. This is the motivation for the second, deeper observation of M87 with XMM-Newton, the results of which are presented in this paper. 

The paper is laid out as follows. In Section 2 we present the data set and data analysis methods. The features seen in the resulting temperature, entropy and pressure maps are described in Section 3. In Section 4 we present more quantitative analyses of these features and possible physical explanations. Our conclusions are summarized in Section 5. We adopt a redshift for M87 of z=0.00436 and a luminosity distance of 16 Mpc \citep{Tonry01}, which yields a scale of 4.65 kpc per arcminute. 
\begin{figure}[t]
\centering
\includegraphics[width=\columnwidth]
{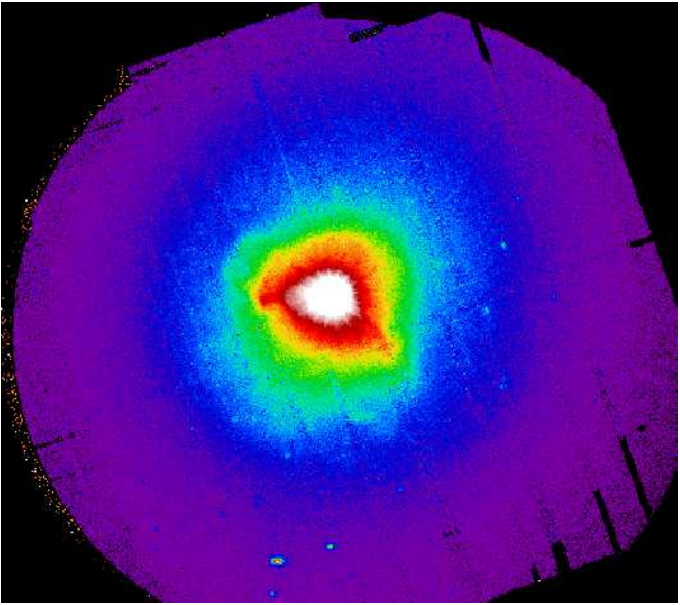}
\caption{M87 surface brightness map, from all three EPIC detectors vignetting corrected and combined, in the 0.5-7.5 keV energy range} \label{sb}
\end{figure}
\section{Observations and data reduction}
M87 was first observed with XMM-Newton in the performance verification (PV) phase on June 19, 2000, for 60 kiloseconds (ksec). A subsequent 109 ksec observation was performed on January 10th, 2005. We will focus in this work primarily on the EPIC data from the second observation, in which the PN detector was operated in the extended full frame window mode while the full frame mode was employed for the MOS detector. The results from the RGS data analysis for the two combined observations can be found in \cite{Werner06}. 

We extracted a lightcurve for each of the three detectors separately and excluded the time
periods in the observation when the count rate deviated from the mean by more than 3$\sigma$
in order to remove flaring from soft protons \citep{Pratt02}. After this cleaning, the net effective exposure
is $\sim$62 ks, $\sim$79.3 ks and $\sim$ 80.5 ks for pn, MOS1 and MOS2, recording in total 13.2, 4.7 and 4.9 million photons
respectively. For data reduction we used the 6.5.0 version of the XMM-Newton Science Analysis System (SAS); the standard analysis methods using this software are described in \cite{Watson01}. A vignetting-corrected flux map of the observation for all three detectors combined in the energy band 0.5-7.5 keV is shown in Figure \ref{sb}. 

For the background subtraction, owing to the extended nature of our source, which covers the entire field of view, we used a collection of blank-sky maps from which point sources have been excised \citep{ReadPonman}. We transposed the blank sky-maps to a position in the sky corresponding to the orientation of XMM-Newton during our observation of M87. The backgrounds were scaled according to the corresponding exposure times for each detector, after having removed the time periods affected by proton flares. The scaling factors in this case are calculated as the ratios between the observation exposure times and the background exposure times. An alternative method for calculating background scaling factors was employed by requiring that in the higher energy region (10-12 keV for MOS, 12-14 keV for PN) the emission in the outskirts of the M87 observation should have the same level as the background, since we expect little or no such high-energy emission from the gas halo around M87. The background scaling factors were in this second case calculated as the ratios between the total number of observation to background counts in the higher energy region specified above, in a 7$^\prime$-9$^\prime$ ring. The values for the scaling factors calculated from the two methods agree to within less than 5\%. This justifies confidence in having chosen a correct scaling.

Where a stable minimum signal-to-noise ratio was needed for our analysis, in particular for determining the temperature, we employed an adaptive binning method based on weighted Voronoi tessellations \citep{Voronoi06}, which is a generalization of the
algorithm presented in \cite{Cappellari03}. The advantage of this algorithm is that it
produces smoothly varying binning shapes that are geometrically unbiased and do not introduce artificially-looking structures, as can be the case for e.g. rectangular bins. Binning the dataset from the second XMM-Newton observation alone to $10^4$ counts per spatial bin in the 0.5-7.5 keV energy range, which should yield temperature values accurate to less than ~5\% for one-temperature model spectral fits, we obtain bins roughly corresponding in size to the extent of the XMM point-spread function (PSF). Therefore, in this paper we only discuss the results from the second observational dataset, which contains enough counts for the purpose of the current analysis. More detailed spectral models and spectral fitting which are the scope of future work will use data from both observations.

Out-of-time events were subtracted from the PN data using the standard SAS prescription.

\subsection{Broad-band fitting}
In order to obtain a first impression of the temperature and abundance distribution of the hot gas
around M87, we restricted the analysis of the surface brightness to four different energy bands selected as follows:
\begin{enumerate}
  \item Low energy band (0.4-0.9 keV)
  \item Fe-L line complex energy band (0.9-1.1 keV)
  \item Intermediate energy band (1.1-2.0 keV)
  \item High energy band (3.0-7.5 keV)
\end{enumerate}
The principle is to compare the trend of fluxes in each pixel in each of the four bands to fluxes expected in these bands from model-spectra for a range of temperatures T=\{0.8,
1.0, 1.5, 1.7, 2.0, 2.2, 2.5, 2.7, 3.0, 3.3, 3.6, 4.0\} keV and metallicities z=\{0.1, 0.3, 0.5,
0.7, 1.0, 1.3\} solar (using the relative abundance values of \cite{Angr}). Subsequently, the best-fitting model was determined using a least-chi-square fit. In all models, the absorbing column density n$_{\rm{H}}$ was set to 1.5$\times10^{20}\:\rm{cm}^{-2}$ \citep{Lieu96}, and the redshift to 0.00436.

We merged the two MOS detectors, which have very similar spectral properties, and produced background-corrected flux maps in each of the four selected bands for the PN and combined MOS detectors. Each map was adaptively binned using a binning scheme based on the combined raw counts image from the three detectors in the 0.5-7.5 keV band.  We performed one fit with a signal to noise ratio (SNR) of 50 and one with a SNR of 100 which ensured a minimum number of counts of 25...100 per bin in each of the four considered bands and especially the high energy 3.0-7.5 keV band which has the least counts.

This method allowed us to clearly identify the eastern and southwestern arms of M87 as lower temperature
features. Furthermore, a lower temperature component could be detected in the direction of the M87
jet (north-west of the core), almost in continuation of each of the arms, spatially connecting them. However, beyond a radius of around 11 arcmin, both the SN 50 and SN 100 temperature maps became very noisy. This was likely due to the fact that our chosen energy bands were tuned to determine rather low
temperatures accurately; for a good determination of higher temperatures it would be
necessary to sample the higher energy regime (above 3.0 keV) with more bands. The abundance pattern was also rather noisy throughout the produced maps; it seemed to be influenced significantly by the PN gap pattern, and did not show a clear distribution relating to the X-ray arms. This was likely due to the fact that our energy bands are very wide, leading to a washing-out of any emission lines with the exception of the Fe-L complex, which is not sufficient for an accurate abundance determination.

In summary, this method was a very fast, computationally cheap way to
gain insight into the spatial distribution of temperatures in the observation. However,
the result seemed to be rather sensitive to the choice of the energy bands, and the insufficient sampling of the higher-energy regime caused a noisy fit in the hotter regions of the gas halo. Moreover, it was an insensitive method for obtaining metallicity maps. 

\subsection{Spectral analysis}

A more reliable method for creating temperature maps is a
comprehensive fit to the spectra from each bin using the XSPEC code, taking into consideration a large number of energy channels and allowing the temperature and metallicity to vary continuously until
a best fit is found. But, given the large number of spatial bins for this observation, this method can be
computationally very demanding.

Point sources were found using the source detection algorithm implemented in SAS and excised from the observation after a visual check to eliminate spurious detections. The central AGN was not considered a point-source contamination and was thus kept in the dataset throughout the analysis. We
then adaptively binned the combined counts image in the 0.5-7.5 keV range using the weighed
Voronoi tessellations method to a target signal-to-noise ratio of 100 and extracted the
spectra of each bin defined in this way. The reason for binning the raw counts map and not the flux map is to obtain larger bins around the gap areas where fewer photons are collected and to ensure an approximately homogeneous accuracy in the temperature determination across the image. A SNR of 100 should give a temperature accuracy better than 5\% (for a one-temperature fit) and a metal abundance accuracy of 10-15\%.

For the spectral analysis we chose the method
described e.g. by \cite{Arnaud01} in which the vignetting correction is
accounted for by adding a weight column to the event file, allowing the use of a single on-axis
ancillary response file (ARF) containing information about the telesope-specific effective area, quantum
efficiency attenuation and filter transmission. The redistribution matrix files on the other hand were computed separately for each bin and each detector. Detector gaps were accounted for by dividing the flux in each bin through the corresponding ratio between the average unvignetted exposure time in the bin and the maximum unvignetted exposure time of the detector.

The spectra in each bin were fitted in XSPEC using a MEKAL one-temperature model. The spectra from all the three detectors were grouped together and fitted simultaneously with the same best-fit model in order to have good statistics. We used the 0.5-7.5 keV range to determine the best fit. However, in the rare bins (approximately 10 bins out of a total of about 1500) where one of the EPIC detectors had an effective
average exposure time of below 10\% of the maximum, the data from that detector was omitted since the number of counts was too low to provide any reliable spectral information. This correction is necessary, as the gaps are not included in the vignetting correction information stored in the weight column.

This method confirms the results obtained from the broad band fitting in that we can clearly
identify the same substructure as noted before: the eastern and southwestern arms of M87 as
lower temperature features, the lower temperature component NW of the nucleus in the direction of the M87 jet, and the nucleus of M87 distinguished as a high-temperature component. We also observe the
positive temperature gradient in the region outside the arms, in agreement with the radial analysis of
\cite{Matsushita02}. The outer parts of the gas halo are significantly less noisy than
derived from the broad band fitting. However, at smaller radii the values of the temperatures determined with the two methods agree very well. Our temperature map is also in good agreement with the temperature map obtained from the previous XMM-Newton observation of M87, computed following the method of \cite{Churazov96} and presented in \cite{Forman05}.

\begin{figure}
\begin{center}
\includegraphics[width=\columnwidth]{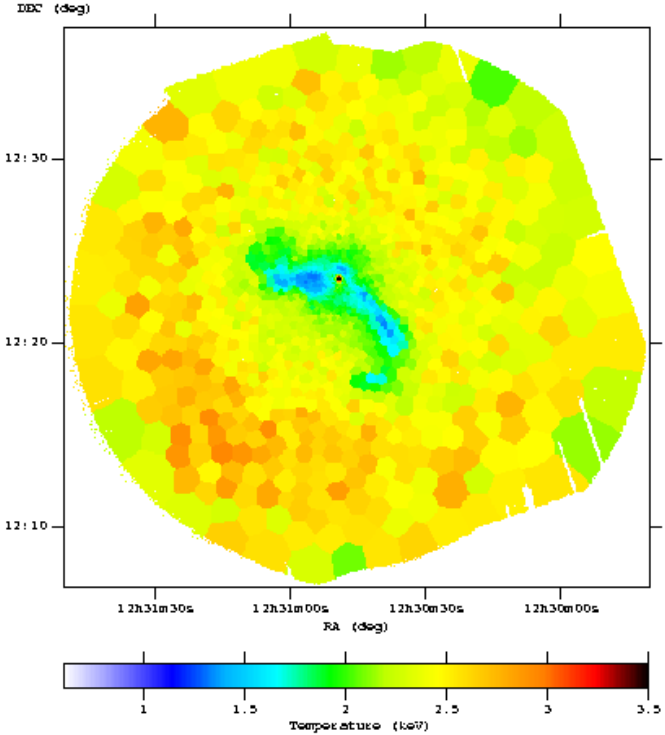}\\
  \caption{Temperature map obtained from spectral analysis, using a binning to SNR of 100.}\label{T_XSPEC}
  \end{center}
\end{figure}
\begin{figure}
\begin{center}
\includegraphics[width=\columnwidth]{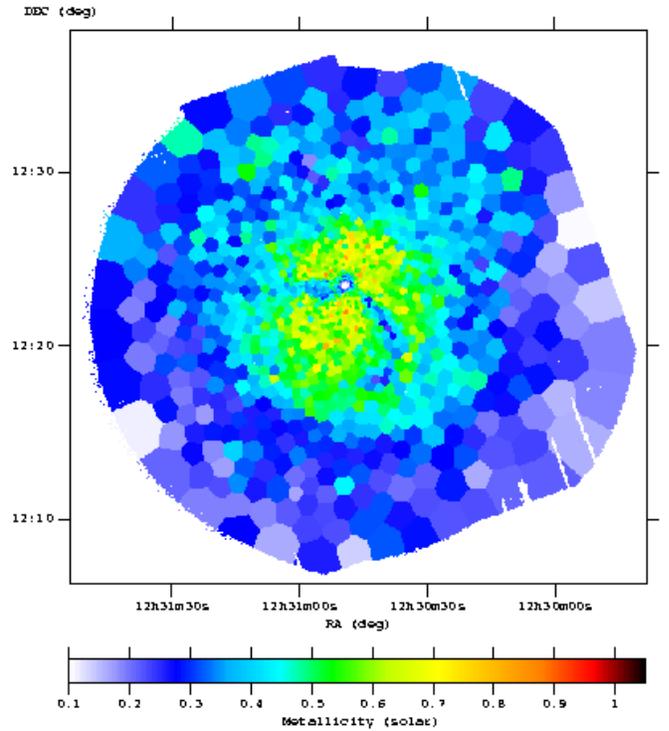}\\
  \caption{Abundance map obtained from spectral analysis, using a binning to SNR of 100.}\label{Fe_XSPEC}
  \end{center}
\end{figure}

The abundance pattern is much better defined with this method compared to the broad-band fitting,
however some noise can still be seen. We note that the statistics in each bin were not sufficient to determine individual element abundances, therefore the abundance ratios were assumed to be solar \citep{Angr}. Detailed radial profiles of the individual elements \citep{Matsushita03} however show that this is not generally the case. Given our signal-to-noise ratios, the obtained abundance values are most likely driven mainly by the iron content whose L-line complex is the most prominent feature in the spectrum. We find that the abundance in the arms is in general lower than in the surrounding medium (Figure \ref{Fe_XSPEC}), which is most likely due to the fact that we are using a single-temperature fit in a region which has been shown to have at least two components with different temperatures (\cite{Belsole01}, \cite{Matsushita02}, \cite{Molendi02}). As these
authors have already noted, a one-temperature fit severely underestimates the
metallicity in this case. Moreover, by creating a map (not shown) of the absorbing hydrogen column
density ($\rm{n_H}$) which was left free in the fit, we find elevated values in the regions corresponding
to the arms, which may also show that a single-temperature fit is not sufficient to appropriately model the data. We performed a second fit fixing the absorbing hydrogen column density to 2.0$\times 10^{20}\rm{cm}^{-2}$ which did not significantly change the temperature map. Therefore, any uncertainty in the $\rm{n_H}$ seems to have little effect on the temperature results; in the following we will base our discussion on the initial temperature results obtained with an unconstrained $\rm{n_H}$ in the fit.

\subsection{Pressure and entropy maps}

We used the temperature and spectrum normalization in each bin to determine a quasi-deprojected measure of the pressure and entropy, which we calculated as $p = n_ekT$ and $S = kTn_e^{-2/3}$. The entropy as defined here is somewhat different from the thermodynamic quantity defined as $S\sim (3/2)k \rm{ln}(T\rho^{-2/3})$ but is nevertheless a measure of adiabaticity and has become commonly used in cluster astrophysics. 

The electron density $n_e$ is determined from the spectrum normalization, which is defined and implemented in XSPEC as $K=10^{-14}/[4\pi D_A^2(1+z)^2]\int n_e^2dV$, from which follows
$$n_e\propto \sqrt{\frac{K}{V}}D_A(1+z)$$ where $D_A$ is the angular diameter distance and z is the redshift. As can be seen from the above formula, we need to also calculate the volume V along the line of sight corresponding to each bin. This was determined as $V\approx(4/3)D_A^3\Omega(\theta_{out}^2-\theta_{in}^2)^{1/2}$, where $\Omega$ is the solid angle subtended by the bin and $\theta_{in}$, $\theta_{out}$ are the angles corresponding to the smallest and respectively largest distances between any of the bin pixels and the M87 nucleus \citep{Henry04}. Thus, this method takes into account an approximate estimation of the three-dimensional extent of each bin, but assumes a constant temperature along the line of sight, and can therefore be viewed as a quasi deprojection. Since most of the emission in each bin originates from the densest gas which is found at the smallest effective 3D radii, the maps can be interpreted as measures of the pressure and entropy in a two dimensional slice through the middle of the cluster, perpendicular to the line of sight. 

\begin{figure*}[h]
\centering
\begin{tabular}{ccc}
\begin{minipage}{3.25in}
\centering
\includegraphics[width=\columnwidth]
{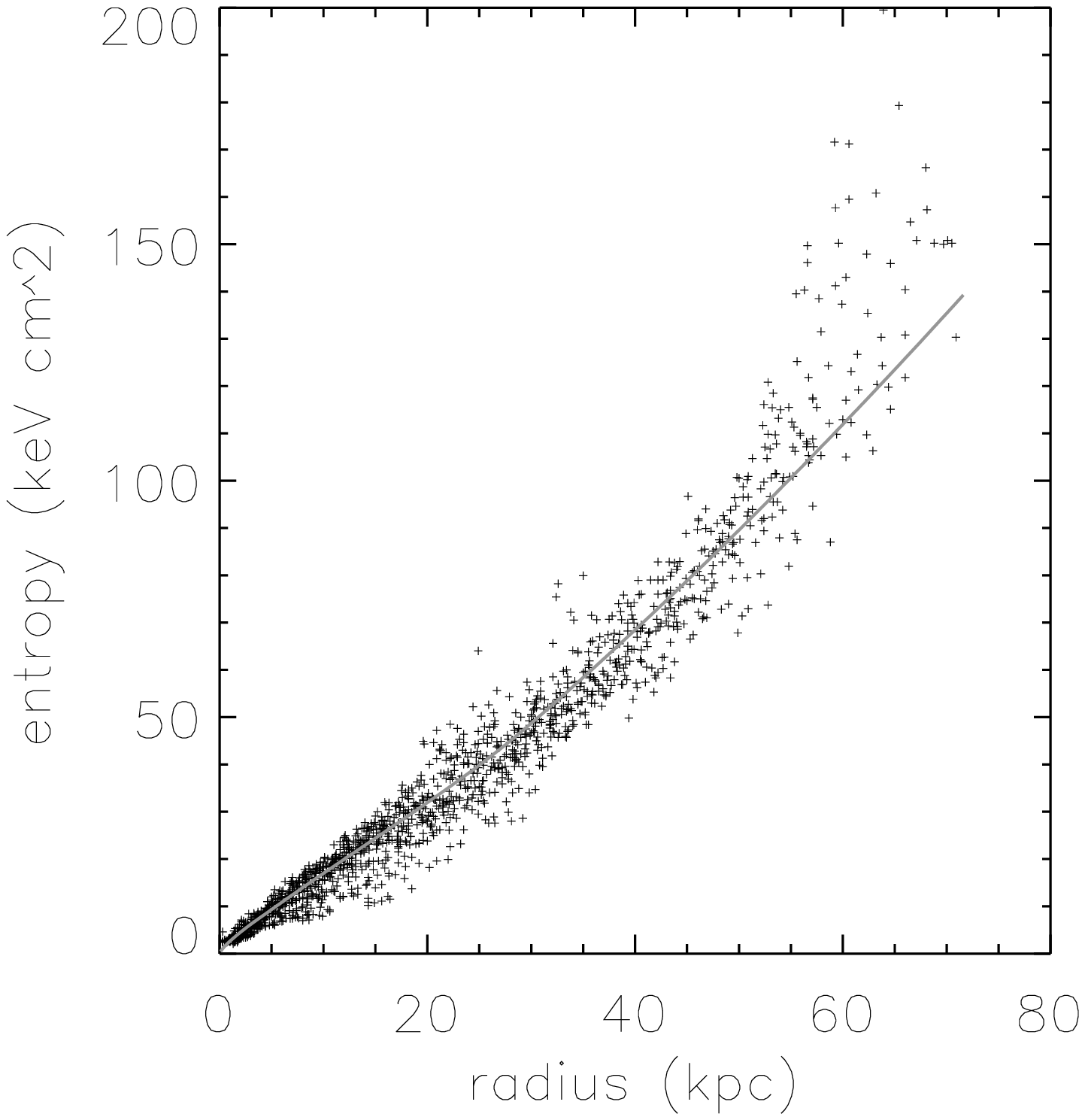}
\end{minipage}
&
\begin{minipage}{3.25in}
\centering
\includegraphics[width=\columnwidth]
{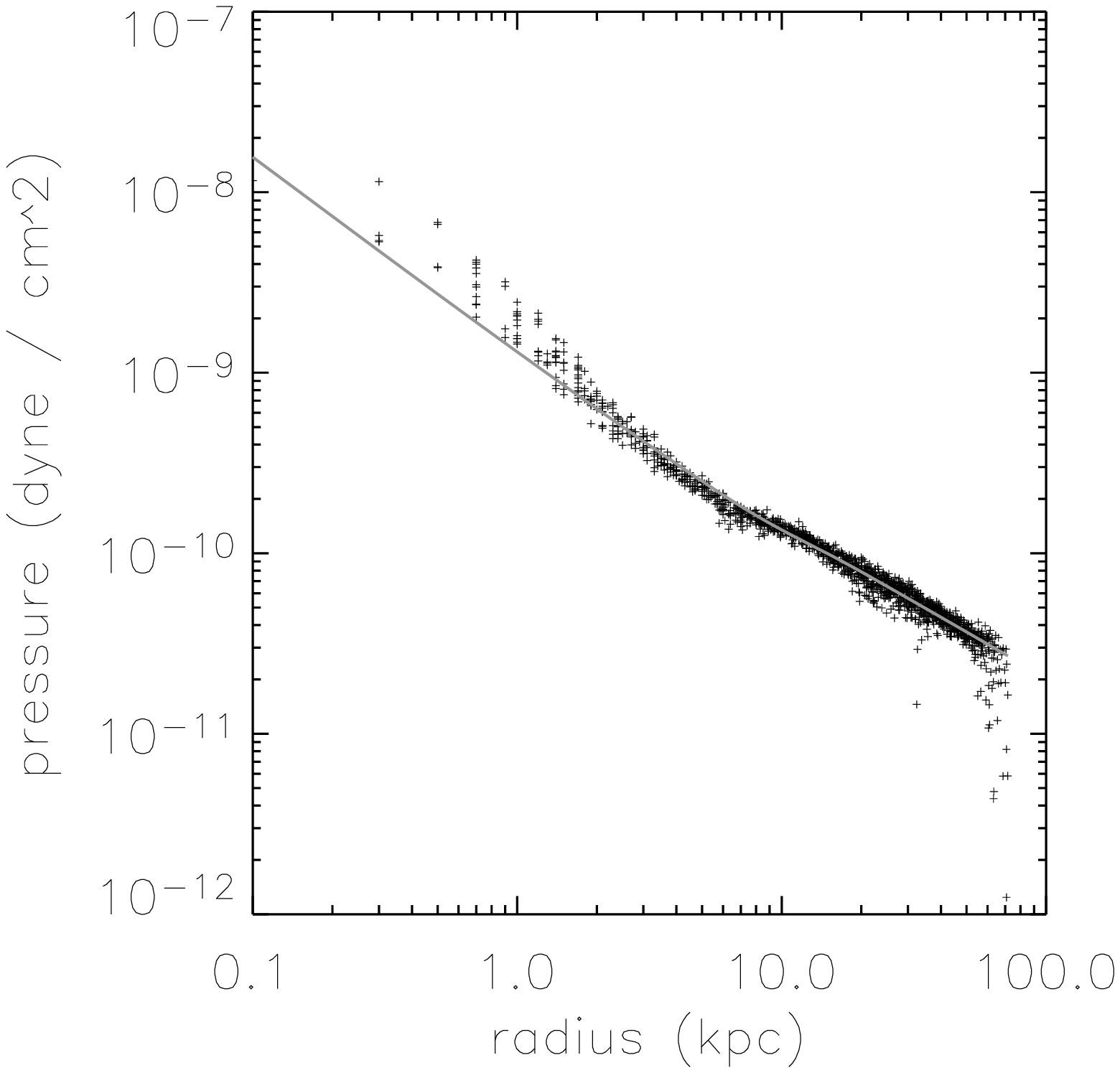}
\end{minipage}
\\
(a) Entropy profile and fitted smooth, non-parametric radial model & 

(b) Pressure profile and fitted smooth, non-parametric radial model

\\
\end{tabular}
\caption{Data points and corresponding smooth radial model by which the entropy and pressure maps were divided in order to reveal small deviations from spherical symmetry} \label{S_p_fit}
\end{figure*}

\begin{figure}
\begin{center}
\includegraphics[width=\columnwidth]{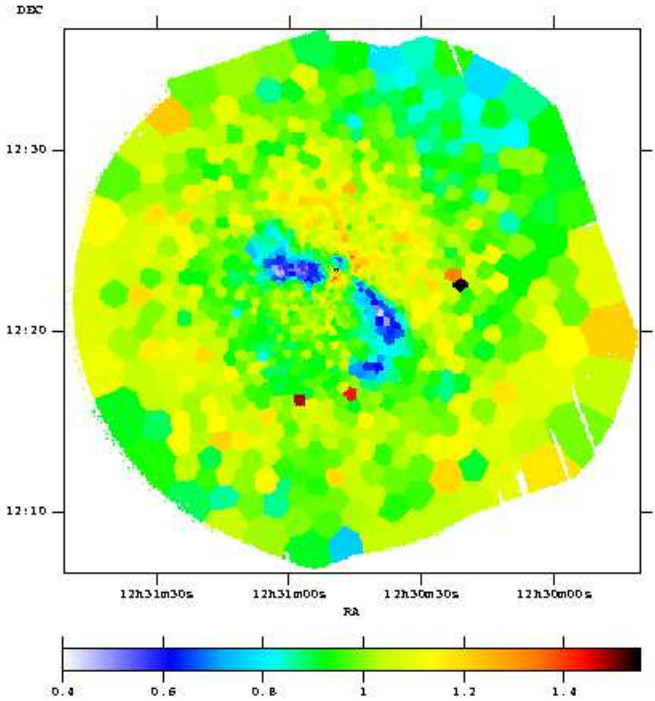}\\
  \caption{Entropy deviations from a smooth radially symmetric model. The E and SW arms, as well as the low-entropy feature close to the NW of the core are seen in dark blue. The entropy edge to the SW is marked by the sharp green to yellow transition. A NW/SE asymmetry is easily seen in the map.}\label{entropy}
  \end{center}
\end{figure}

We fitted the radial pressure and entropy profiles using non-parametric, locally weighted, linear regression smoothing. The data points and the corresponding smooth radial model are overplotted in Figure \ref{S_p_fit}. The pressure and entropy maps were divided by the resulting radial model in order to reveal small-scale fluctuations (Figures \ref{entropy} and \ref{pressure}). We can identify several features pointed out by \cite{Forman05}, such as the 3$^\prime$ (14 kpc) ring of enhanced emission corresponding to an increase in the pressure. This confirms the association of this substructure with a shock, while the NW/SE brightness enhancements seen with Chandra \citep{Forman05} appear as low-entropy features. The X-ray arms are clearly low-entropy structures, which also seems to confirm the hypothesis that the gas in these regions was uplifted from more central regions where we expect the smallest entropy in hydrostatic equilibrium conditions. The next section presents a more detailed analysis of these features.

\section{Substructure in the M87 gas halo}
\subsection{Temperature and Entropy}

The temperature and entropy maps both show similar substructure details, so that we chose to discuss them in parallel. By far the most striking features in both maps are the E and SW arms characterized by lower temperature and lower entropy with respect to the surroundings. The description of the thermal structure of the arms will be the subject of an upcoming paper in which a more detailed spectral analysis will be employed in order to understand the complex multi-temperature structure that several authors have already found in these regions (\cite{Molendi02}, \cite{Matsushita02}, \cite{Belsole01}). We point out that in the temperature and entropy maps both arms are seen to curve clockwise, thereby connecting to the larger-scale radio halos north and south of the nucleus. This was already known for the SW arm (see for example \cite{Forman05}), but had not previously been observed for the eastern arm.

Apart from the E and SW arms, another low-temperature and low-entropy feature can be identified close to the nucleus towards the NW, almost at the intersection of the lines defining the general directions of the arms. This region corresponds to high X-ray luminosity and low entropy, implying a locally high gas density. Additionally, enhanced H$\alpha$ emission has been observed here \citep{SparksDonahue04}, which makes this region an attractive candidate for hosting a classical cooling flow and associated mass deposition. However, detailed spectral analysis of the region, as described in the next section, does not provide evidence for cooling to the lowest X-ray detectable temperatures at the rates predicted in the absence of any heating sources.

Another feature found in both maps is an edge to the SE at a radius of roughly 6$^\prime$. As one moves outwards across this edge, both the temperature and entropy increase, therefore the properties at the jump are more consistent with the cold-front interpretation of \cite{Markevitch01} than with the characteristics of a shock. A more detailed discussion of this is presented in the next section.

Finally, a feature easily seen in the entropy map but which is not readily apparent in the temperature map is a NW/SE asymmetry. Within a radius of 6$^\prime$ the entropy values to the NW are clearly more elevated than in the SE. The edge to the NW where the entropy decreases corresponds spatially very well to the edge of the NW large-scale radio bubble. Possible physical explanations for this are also presented in the next section.

\subsection{Pressure}
A map of the pressure deviations from a radially symmetric, smooth model shows three noteworthy features. The first of these is a relative pressure increase towards the NW and SE. Since we subtracted a radially symmetric model, this would suggest an overall ellipticity of the pressure distribution and consequently of the underlying dark matter distribution under the assumption of hydrostatic equilibrium. Secondly, in the direction of the E and SW X-ray arms (coinciding with the prominent radio lobes and orthogonal to the regions of enhanced pressure), we find a pressure decrease. Thirdly, there is a ring of enhanced pressure with a radius of roughly 3$^\prime$, coinciding with the position of the weak shock proposed by \cite{Forman05}.
\begin{figure}
\begin{center}
\includegraphics[width=\columnwidth]{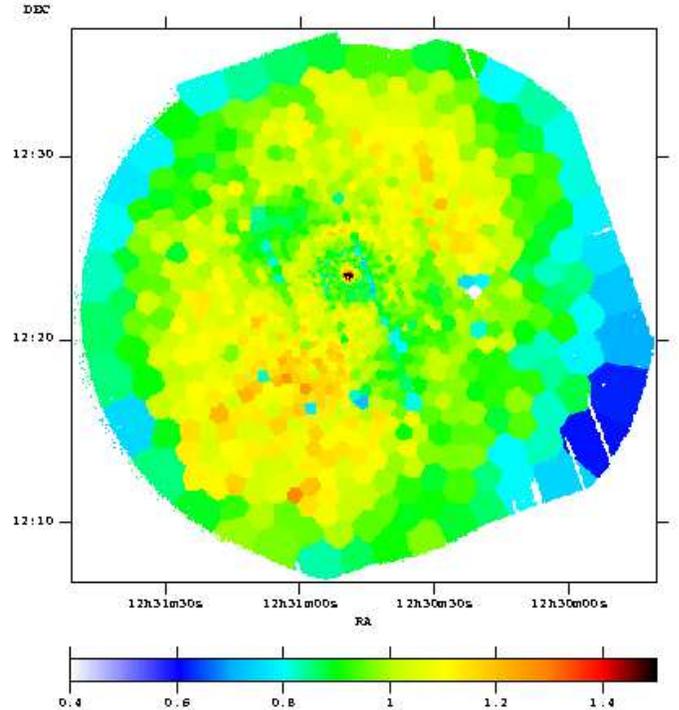}\\
  \caption{Pressure deviations from a smooth radially symmetric model. A relative pressure increase towards the SE and NW suggests an overall ellipticity in the pressure. A pressure decrease is found in the direction of the E and SW arms, which rise perpendicular to the SE/NW ellipticity long axis. A ring of enhanced pressure with a radius of roughly 3$^\prime$ is also seen.}\label{pressure}
  \end{center}
\end{figure}

We tried to confirm and quantify the ellipticity in the initial pressure map by fitting to it elliptical isocontours with the "ellipse" task in IRAF. The position angle and ellipticity were independently fitted for each pressure contour and the model need not be smooth in intensity as a function of the semi-major axis. We masked the regions corresponding to the radio lobes from the fit, since here it is likely that the pressure deviates from an elliptical model. 
\begin{figure}
\begin{center}
\includegraphics[width=\columnwidth]{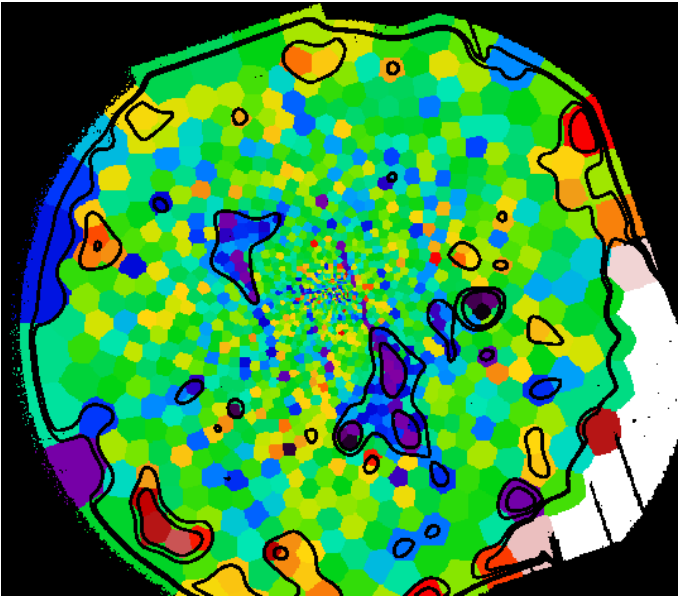}\\
  \caption{Pressure deviations from an elliptical model (ratio). Contour levels are drawn at 0.9, 0.95, 1.05 and 1.1. Two regions roughly corresponding to the end of the E arm and to the SW arm are seen to have lower than average values, indicating the possible presence of nonthermal pressure support.}\label{pressure_ellip}
  \end{center}
\end{figure}
Dividing the initial pressure map by the fitted model obtained with IRAF, we are able to identify deviations from an elliptical rather than a radially symmetric model (Figure \ref{pressure_ellip}). We find that, with respect to the elliptical model, all substructure seen previously no longer appear, with the exception of the lower pressure in regions that coincide with the radio lobes. Here the decrease in observed pressure with respect to the elliptical model is on average 5-10\%. Assuming an overall pressure balance, this decrease suggests that in the radio lobes there should be an additional contribution to the pressure. This pressure source may be provided by relativistic electrons injected by the AGN. The ring at 3$^\prime$ is no longer visible since we did not require the pressure intensity in the model to be smooth with increasing semi-major axis, while the higher pressures NW and SE of the nucleus which appeared as deviations from a radially symmetric model can be entirely explained by an elliptical pressure distribution.

\section{Discussion}
\subsection{Absence of a classical cooling-flow on smaller scale}
 \begin{figure}
\includegraphics[height=\columnwidth,angle=-90]{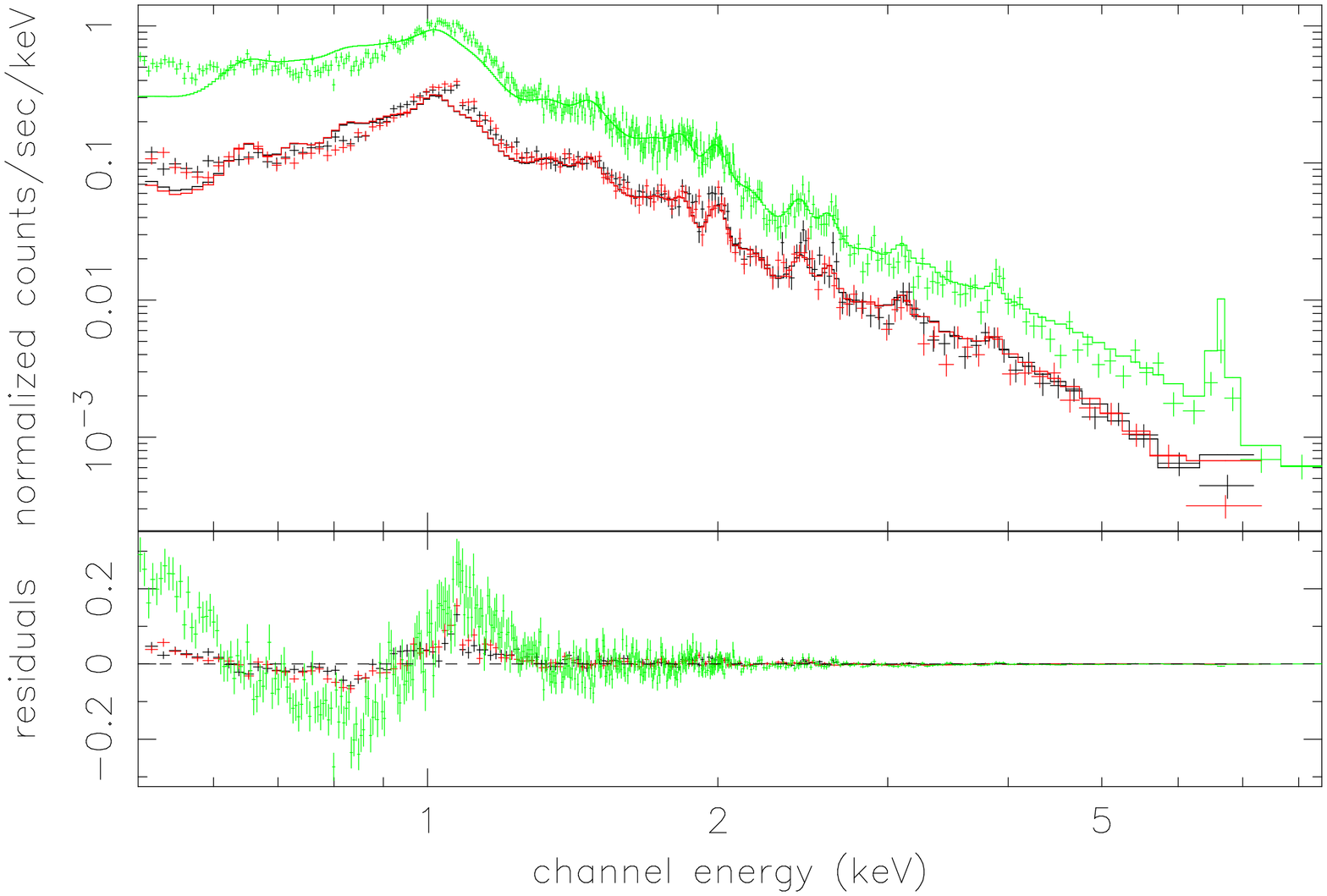}\\
  \caption{Data and best-fit vmcflow model with fixed low-temperature cutoff for the small-scale low-temperature feature NW of the core. The disagreement of the model with the data is easily visible, indicating the absence of a classical cooling flow.}\label{vmcflow}
  \vspace{-0.25cm}
  \includegraphics[height=\columnwidth,angle=-90]{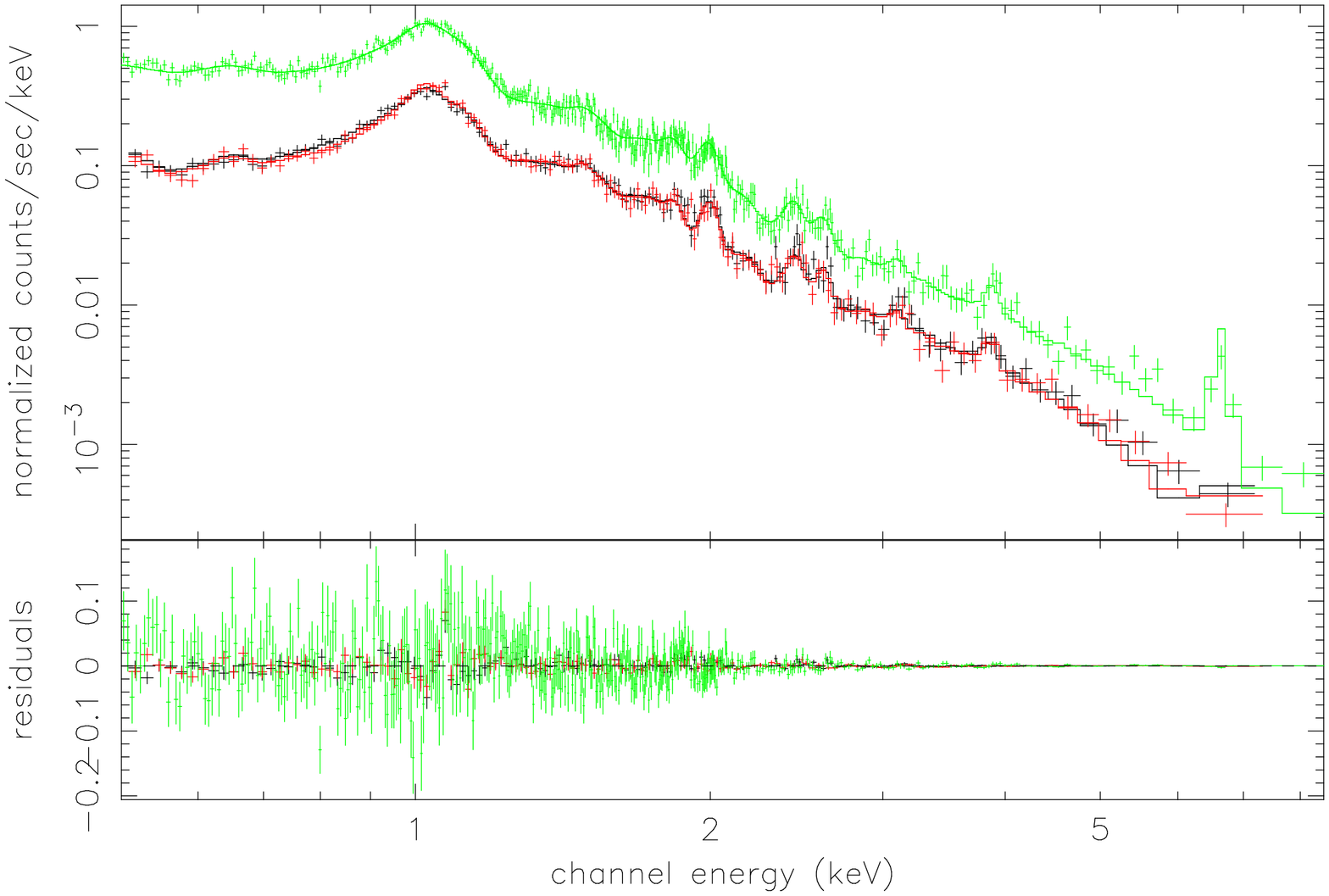}\\
  \caption{Data and best-fit two-temperature vmekal model for the small-scale low-temperature feature NW of the core. This model is clearly in better agreement with the data than the classical cooling flow model.}\label{vmc_noflow}
\end{figure}
While a large-scale cooling flow has been ruled out, cooling and condensation could still happen locally, since it is not easy to exactly balance heating and cooling everywhere throughout the cooling core region. A prime candidate target to look for such a local cooling flow region is the small low-entropy feature NW of the nucleus where also diffuse optical emission lines are observed. 

We extracted a spectrum from this region and fitted it in XSPEC with a vmcflow model fixing the low temperature cutoff at the minimum level available in the model (81 eV) in order to probe the existance of a classical cooling flow spectrum. We find a very poor fit, especially around the Fe-L complex where the data lie below the spectral model between 0.6-1 keV (Figure \ref{vmcflow}). A vmekal+vmcflow model with a fixed low-temperature cutoff at 81 eV provides a good fit for an ambient temperature (vmekal) of 1.508 $\pm$ 0.027 keV and a mass deposition rate of 3.18 $\pm$ 0.16 $\times 10^{-2}$ M$_{\odot}$/year. According to a classical cooling flow analysis done by \cite{Boehringer99}, the mass deposition rate in this region in the absence of a heating source should be 0.1-0.2 M$_{\odot}$/year, which is a factor of about 5 higher than what we find from spectral fitting. A two-temperature vmekal model with T1 = 1.016 $\pm$ 0.022 and T2 = 1.855 $\pm$ 0.030 also describes the data very well. Therefore, although the density in this smaller-scale region is very high and although the region is associated with H$\alpha$ emission, we find that it does not exhibit spectral evidence of a classical cooling flow and sufficient mass deposition rate.
\begin{figure*}
\begin{center}
\includegraphics[width=\textwidth]{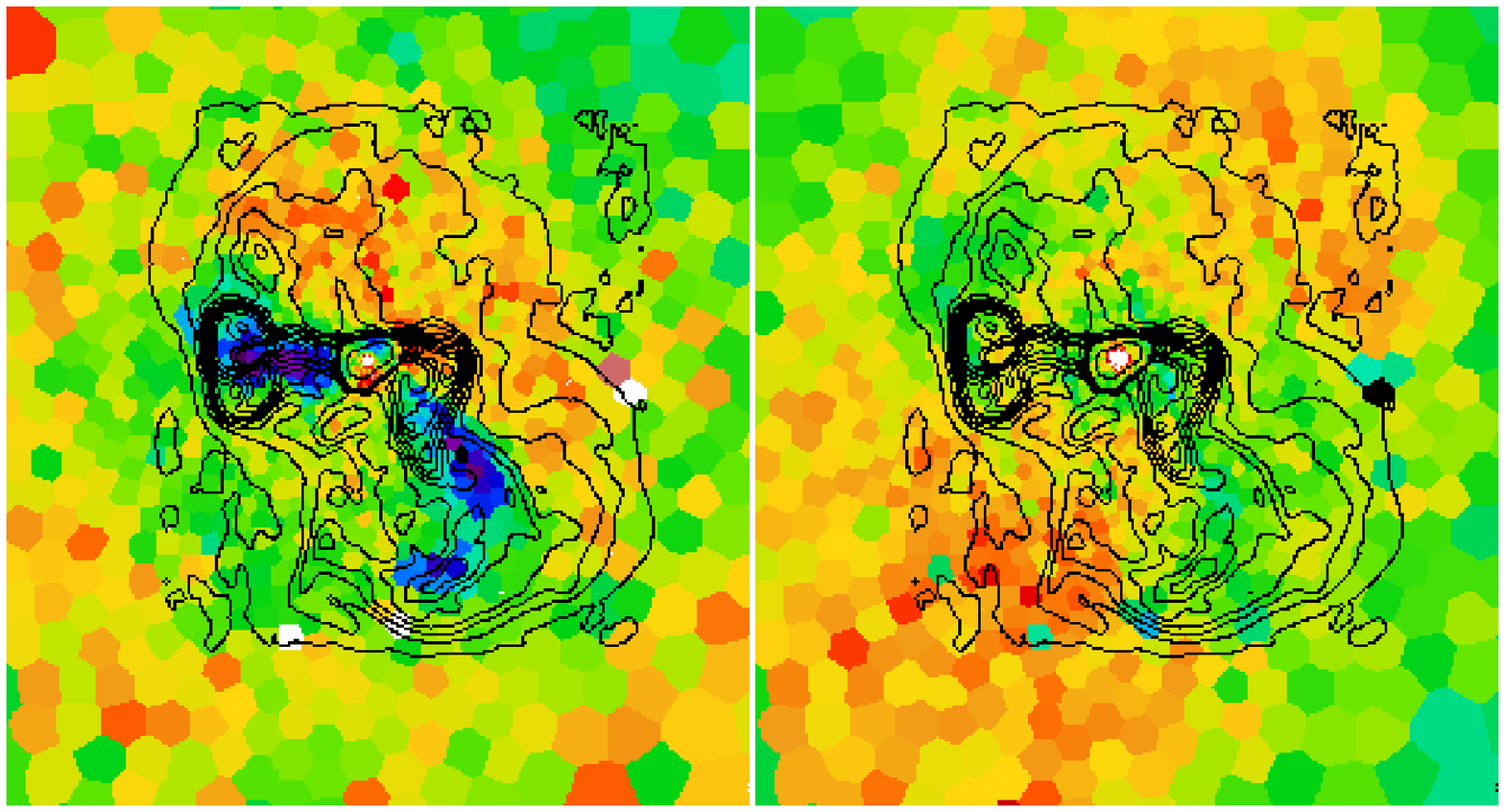}\\
  \caption{Entropy (left) and pressure (right) deviations from a smooth radially symmetric model overlaid with radio contours (90 cm). Radio map kindly provided by F. Owen. The E and SW radiolobes coincide well with the regions of low entropy in the X-ray arms. Also, the edge of the large radiolobe to the north roughly coincides with a NW edge in the entropy map.}\label{radio_overlay}
  \end{center}
\end{figure*}
\subsection{Motion of the galaxy to the NW}
\begin{figure}[h]
\centering
\includegraphics[width=\columnwidth]{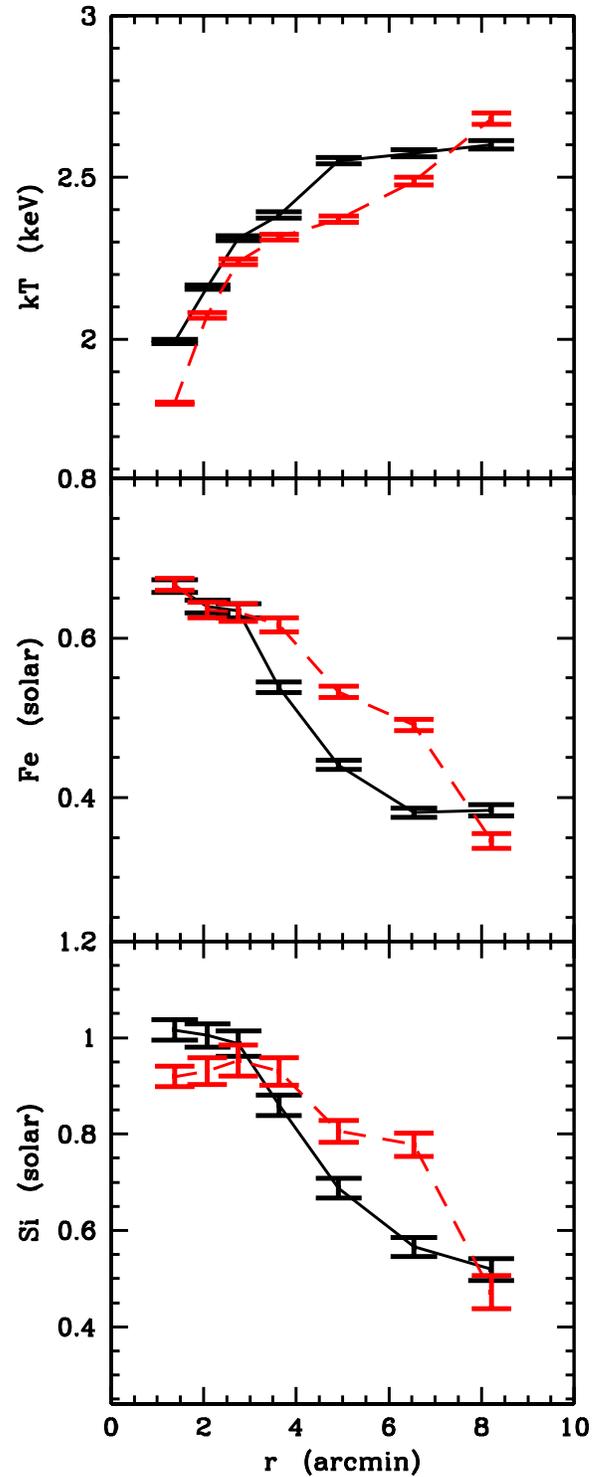}
\caption{Temperature and abundance NW/SE asymmetry. The solid line represents the profile NW of the core, the dashed red line SE of the core.} \label{NW_SE_asymmetry}
\end{figure}

To investigate the NW/SE asymmetry seen in the entropy map, we extracted spectra from concentric ring sectors towards the NW (-25 to 135 degrees counterclockwise of West) and SE (200 to 280 degrees counterclockwise of West). The angles are chosen such that the E and SW arms are excluded. The spectra were fitted with a one-temperature vmekal model, since we expect from the results of \cite{Matsushita02} and \cite{Molendi02} that the gas outside the arm regions is single-phase locally. The results plotted in Figure \ref{NW_SE_asymmetry} show that the radial temperature profile towards the SE lies below the profile towards the NW, while the radial abundance profile reveals higher values in the SE than in the NW.
The simplest phenomenon that explains such a NW/SE temperature and abundance asymmetry is a motion of M87 to the NW through the hot gas halo, or alternatively, a large-scale bulk motion of the hot gas halo towards the SE while the galaxy itself remains in place. In either case, there are two possible mechanisms related to the motion which would generate the observed asymmetry. 

The first possibility is that colder, highly-abundant gas near the center is ram-pressure stripped during the motion. In the case of a large-scale bulk motion of the gas halo, the stripped gas is then carried towards the SE along with the bulk flow, while if the galaxy is moving towards the NW it is left behind in a wake. However, the pressure map is symmetrical and shows no increase to the NW as one would expect if ram-pressure were important. 

The second possibility, which is favoured by our entropy and pressure maps, is that, due to the relative velocity between the nucleus and the gas halo, bubbles emitted initially to the north-west are advected downstream. This effect is important if the relative bulk velocity is at least as high as, and of the order of, the bubble rising velocity and would imply that a large part of the bubbles emitted by the AGN eventually rise to the SE independently of their initial direction of emission. Much more mixing is therefore induced towards the SE, favoring the transport of metals, which explains the relative iron and silicon abundance increase in this direction (Figure \ref{NW_SE_asymmetry} (b) and (c)). Recent simulations \cite{Roediger06} indeed show that bubble-induced metal transport results in elongated abundance profiles along the direction of propagation of the bubbles. The observed NW/SE temperature asymmetry can be explained using this bubble-advection scenario if we consider that the bubble-induced mixing not only favors the transport of metals in the atmosphere but can also bring colder gas from the center out to larger radii. It has been already noted (e.g. \cite{Nulsen02}) that radio bubbles are often surrounded by rims of cold, dense gas which they uplift during their inflation and rising. Therefore it is probable that a preferential propagation of a series of bubbles to one direction may result in lowering the downstream average temperature.

Further indication that bubbles are advected downstream is provided by Chandra data in \cite{Forman06}; the authors describe a series of four consecutive bubbles to the SSE while no counterpart of such structure is seen in the opposite direction. 

Finally, one other observational feature leads us to believe that M87 is moving to the NW. If we consider the analogy of M87 with a typical wide-angle tailed (WAT) radio source, the motion through the intracluster medium may help to explain why the E and SW radio lobes, and the corresponding X-ray arms, do not lie along directions exactly opposite to each other. However, the wide angle between the lobes and especially the radio "ear" seen at the end of the eastern arm suggest that the buoyant velocity is at least comparable to the bulk flow velocity. This agrees qualitatively with our previous statement that the relative bulk velocity should be on the order of the bubble rising velocity.

Let us reconsider the model of ICM heating through radio-jet inflated bubbles. \cite{Birzan04}, among others, show that the bubble enthalpy is in most cases enough to halt the cooling flow. However, the picture drawn by \cite{Nulsen02} for Hydra A and in this work also for M87 is much more complex. While the AGN does inject energy into the ICM in the form of bubble enthalpy, bubble formation is likely often associated with the uplift and mixing of cool gas from the center. A well known example for the uplift are the X-ray arms of M87 \citep{Churazov01}. In this paper we also find for the first time evidence of this phenomenon at smaller temperature and density contrasts with respect to the ambient medium within the same galaxy, contrasts which could therefore not have been observed without the high-quality statistics of the dataset. This suggests that cool-gas uplifting by buoyant bubbles may be more wide-spread than previously thought, and has not been detected so far solely owing to insufficient statistics. Whether the uplifted cool negatively buoyant gas will fall back towards the cluster center thermalizing its potential energy \citep{Nulsen02}, or whether it is heated by thermal conduction once it reaches larger radii, is a matter for future models; however, it seems that bubbles do generally transport cool gas out of the center.

\subsection{Possible core oscillations}
While the abundance and temperature profiles of the X-ray gas relatively close to M87 (within the inner $\sim$ 6$^\prime$) are well explained by a simple relative motion between the gas and the galaxy, a look at the features seen further towards the outskirts reveals a more complex nature of the ongoing physical phenomenon. The model-divided entropy map reveals an edge at about 6$^\prime$ SE of the core which is suggestive of a cold front, with the pressure across the front staying roughly constant and the temperature and entropy being lower on the more X-ray luminous side. This feature corresponds in the metallicity map (Figure \ref{Fe_XSPEC}) to a visible sharp edge in the abundance distribution, emphasizing the presence of a contact discontinuity here. Another relatively sharp edge in the metallicity can also be seen at about 3$^\prime$ NW of the core, although it is not associated with any visible entropy or temperature features. Moreover, also NW of the core, we find a region of lower entropy beyond 6$^\prime$. 

The opposite and staggered placement of these low-entropy regions and abundance edges suggests that the current relative motion of the gas and the galaxy, proposed in the previous section, may be part of a longer process of oscillatory motions along the NW/SE direction. 
\cite{Churazov03} observed a similar E/W asymmetry and a set of sharp edges in entropy, temperature and surface brightness in the Perseus cluster and performed simulations which suggest that minor mergers may trigger oscillations of the cluster gas, thus generating multiple sharp edges on opposite sides of the central galaxy along the direction of the merger. Similar opposite and staggered features generated by oscillations of the intracluster gas are seen also in numerical simulations by \cite{Tittley05} and \cite{Ascasibar06}. 

Further evidence of a possible oscillatory motion of the gas in the M87 halo comes from an analysis of the pressure map. Fitting an elliptical model to the raw pressure map as described in Section 4.5 and allowing the centers of the elliptical isobars to vary, we find that these centers systematically shift towards the SE for higher values of the semimajor axis. Thus the galaxy potential located at smaller scales is not exactly centered with the large-scale halo potential, which would lead to the oscillation suggested here. In a more detailed future analysis, this can be used to reveal information about the amplitude, energy, and time-scale of these oscillations, which may help in understanding how this phenomenon contributes over time to the heating vs. cooling balance, and how it competes with other proposed methods for energy input into the ICM. 

\subsection{Large-scale correlation of the radio emission and entropy map}

One of the striking large-scale features in Figure \ref{radio_overlay} is the spatial coincidence of the northern outermost radio contour-edge and a decrease in the entropy by 10-15\%, which cannot be explained either by a directed motion of M87 to the NW, or by the possible core oscillations. \cite{OEK00} note that the radio contour-edge represents a sharp decrease in radio intensity over a range of wavelengths, therefore indicating interaction between the radio-loud plasma and the hot gas halo.

\begin{figure}
\begin{center}
\includegraphics[width=\columnwidth]{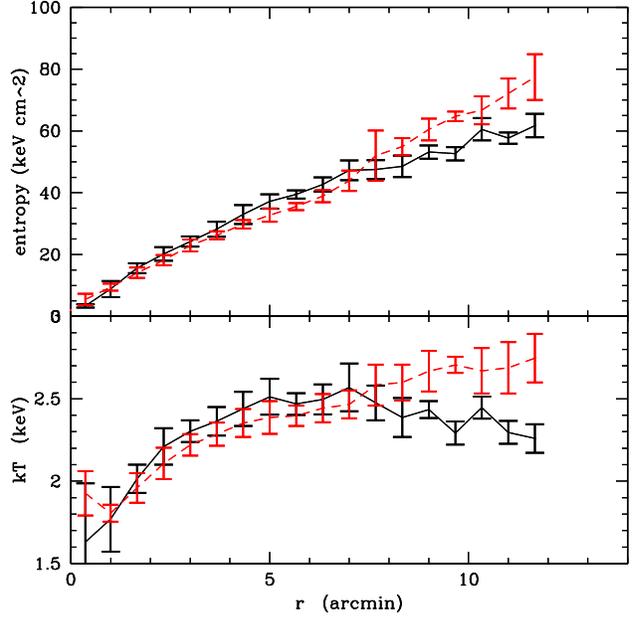}\\
  \caption{Entropy and temperature radial profiles, calculated as an average of the values in the bins whose center falls inside circular annuli sectors with an opening angle of 90 degrees NW and SE of the core. The errorbars represent the corresponding standard deviation in each annulus sector.}\label{entropy_radiojump}
  \end{center}
\end{figure}

To illustrate the entropy variation across the radio edge, we averaged the entropy in circular annuli sectors with an opening angle of 90 degrees in the NW and SE quadrant respectively. We compare the radial profiles of the entropy in these two quadrants in Figure \ref{entropy_radiojump}, which also shows radial trends of the temperature in the corresponding regions. Up to a radius of roughly 6$^\prime$-7$^\prime$ the entropy in the NW quadrant is higher than in the SE quadrant. Beyond this radius, which corresponds roughly to the radio edge, the entropy and temperature NW of the core show a much shallower increase with radius than in the SE, the temperature displaying even a small negative gradient. 
The spatial coincidence between the north radio edge and the point where the entropy and temperature gradients NW of the core decrease suggests a link between the injection of radio-loud plasma into the ICM and the processes leading to entropy generation and gas heating. Such a link is at the basis of the feedback models in which mechanical input, usually by a radio source, offsets the cooling flow, heating the gas at the cluster center and flattening the central entropy profiles. The exact details of the entropy-generation mechanism remain unclear.   

\subsection{Ellipticity of the underlying dark matter potential}
As mentioned above, we fitted the raw pressure map with an elliptical model in IRAF in order to quantify the eccentricity and position angle-dependency with semi-major axis. The results of the obtained fit show larger variations in the central part but outside 3$^\prime$ a rather constant position angle and ellipticity are reached. The position angle values outside 3$^\prime$ range between 140 and 150 degrees, measured counterclockwise of north. These values are somewhat smaller than, but comparable to, results from the analysis of an optical image of M87 by \cite{Carter78}, who obtained values for the elliptical isophote-fit of 160 $\pm$ 5 degrees for the position angle. Our results agree also with the work of \cite{Boehringer97} on ROSAT PSPC data, giving an ellipticity of the X-ray surface brightness with a position angle of 158 $\pm$ 10 degrees. Outside of 3$^\prime$, the ellipticity varies generally between 0.08 and 0.17, higher values of 0.13-0.17 being found in the regions from which the radio lobes had been cut out. This is in agreement with \cite{Boehringer97}, who found an ellipticity of 0.1 to 0.16. The ellipticity in the X-rays thus seems to be significantly smaller than in the optical regime, where it ranges from 0.3-0.5. This can be explained by the fact that anisotropic velocity dispersions of the stars can sustain a higher ellipticity of the galaxy in the optical domain while the isotropic nature of the X-ray emitting gas halo makes it more round. Surprisingly, despite the evidence of core oscillations seen in the entropy map, the position angles of the X-ray isobars and the optical isophotes agree very well, and the optical ellipticity is as expected roughly 3 times higher than the X-ray value (see also \cite{Buote96}). Therefore the deviations of the hot gas atmosphere from hydrostatic equilibrium cannot be very large outside the arm regions.

A further look at the position angle values found in the fit reveals the fact that the radio lobes rise in directions orthogonal to the semimajor axis of the elliptical dark matter profile, thereby following the steepest dark-matter potential gradient as one would expect. 

\subsection{Heating of the ICM by weak shocks}
\begin{figure}[h]
\centering
\includegraphics[bb = 18 144 477 718, width=\columnwidth]{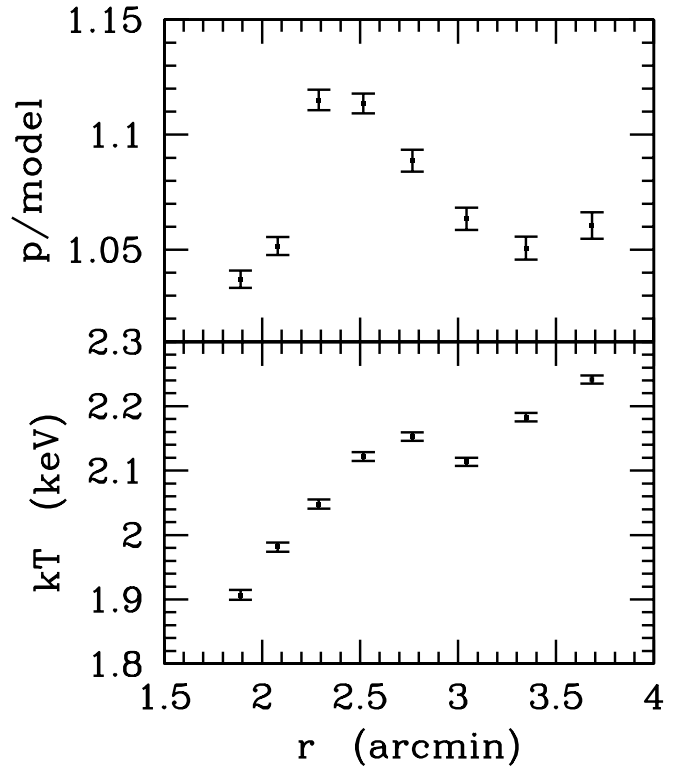}
\caption{Projected temperature and relative pressure jumps around the 3$^\prime$ shock, corresponding to Mach numbers of 1.05 and 1.04 respectively.} \label{t_p_jump}
\end{figure}

As seen in Figure \ref{pressure}, we are able to spectroscopically confirm the 3$^\prime$ shock seen by \cite{Forman05}, \cite{Forman06} with Chandra. Weak shocks are becoming widely accepted as candidates for heating mechanisms of the X-ray gas, therefore a detailed understanding of their properties is required. To describe the M87 3$^\prime$ shock more quantitatively, we used the elliptical isobars from the fit described above to define eight spectral extraction regions with semi-major axes between 2 and 4 arcminutes. The temperature and spectrum normalization corresponding to the regions contained between each two consecutive elliptical isobars in the chosen semi-major axis range were determined using a vmekal one-temperature spectral fit. The corresponding pressure was then determined for each region as described in Section 2.3. The results are plotted in Figure \ref{t_p_jump}. In the plot, the pressure was divided by a smooth model to emphasize the jump.

The temperature shows a jump of $\sim$0.1 keV, which relative to the overall temperature of $\sim$2 keV, represents $\sim$5\%. Using the Rankine-Hugoniot jump conditions assuming $\gamma=5/3$, this amounts to a Mach number of 1.05. The pressure jump amounts to 8-10\% with respect to the smooth model, which implies a Mach number of about 1.04, in agreement with the Mach number we derived from the temperature jump. This confirms the supersonic nature of the weak shock at 3$^\prime$, and seems consistent with the Mach number M$\approx$1.2 calculated from deprojection analysis of Chandra data \citep{Forman06} considering that the Mach numbers derived from this analysis are lower limits to the real values due to projection effects and the smoothing effect of the XMM PSF. Accurate spectral deprojection analysis is the scope of upcoming work.

\section{Summary and conclusions}
We present the results from a 109 ksec XMM-Newton observation of the hot gas halo surrounding the giant elliptical galaxy M87. We use two methods, namely broad-band fitting and full spectroscopy, to create spatially resolved temperature maps from the data, and present the advantages and disadvantages of each method. The results of the spectroscopic analysis are used to produce entropy and pressure maps. We describe and discuss the features seen in these maps, the most important of which being the cool E and SW X-ray arms, which coincide with powerful radio lobes, a weak shock ring with a radius of 3$^\prime$, an overall ellipticity in the pressure map and a NW/SE asymmetry in the entropy map. For the 3$^\prime$ weak shock we find jumps in both the pressure and temperature corresponding to Mach numbers of 1.04 and 1.05 respectively, which we interpret as lower limits to the real Mach number values. The pressure map shows an overall ellipticity of between 0.08 and 0.17, with position angles in good agreement with the optical data. Under the assumption of hydrostatic equilibrium, the ellipticity in the pressure map implies an elliptical underlying dark matter potential. Furthermore, we find that the X-ray arms and associated radio lobes rise roughly perpendicular to the semimajor axis of the pressure ellipticity following the steepest gradient of the dark matter potential.

The NW/SE entropy asymmetry is indicative of the motion of M87 through the surrounding hot gas, associated with downstream advection of the bubbles injected by the AGN. We conclude that bubble-induced cold gas mixing from the center may appear on a wide variety of scales and should be considered into the cooling-flow energy balance calculations. NW and SE edges in the entropy map moreover suggest that the possible motion of M87 to the NW is part of a longer cycle of core oscillations.

\begin{acknowledgements}
We would like to thank W. Forman, P. Nulsen, E. Churazov and G. W. Pratt for helpful discussion. We acknowledge the support by the DFG grant BR 2026/3 within the Priority Programme "Witnesses of Cosmic History". The XMM-Newton project is an ESA Science Mission with instruments and contributions directly funded by ESA Member States and the USA (NASA). The XMM-Newton project is supported by the Bundesministerium fuer Wirtschaft und Technologie/Deutsches Zentrum fuer Luft- und Raumfahrt (BMWI/DLR, FKZ 50 OX0001), the Max-Planck Society and the Heidenhain-Stiftung, and also by
PPARC, CEA, CNES, and ASI. AF acknowledges support from BMBF/DLR under grant 50 OR 0207 and MPG. This project is partially supported by a NASA grant NNG05GH40G to UMBC. AS also thanks Steven Diehl and Thomas Statler for making their Weighted Voronoi binning algorithm publicly available.
\end{acknowledgements}

\bibliographystyle{aa}
\bibliography{6650bibliography}

\end{document}